\documentclass[cup7b]{cupbook}
\usepackage{graphicx,calc,units}
\usepackage{amsmath,amssymb,amsfonts,bm}

\newcommand\ve[1]{\bm{\mathbf{#1}}}

\begin{document}

\maketitle % this \inputs \jobname.ttl if available
           % Otherwise its use gives blank pages and allows second
           % \maketitle to act on Chapter authors.

\author[J. Collins, A. Perez and D. Sudarsky]
{John  Collins\\  Physics Department, Pennsylvania State University\\
University Park, PA  16802 USA
\and
 Alejandro Perez\\
Centre de Physique Th\'eorique, Campus de Luminy,
13288 Marseille, France
 \and  Daniel Sudarsky\\ Instituto de Ciencias Nucleares,  Universidad Aut\'onoma de M\'exico\\
  A. P. 70-543, M\'exico D.F. 04510, M\'exico\\
 }

\maketitle

\chapter[Lorentz Invariance \& Quantum Gravity Phenomenology]{Lorentz Invariance Violation and its Role in Quantum Gravity Phenomenology} % chap 3

\vspace*{1cm}
\begin{abstract}
  The notion that gravitation might lead to a breakdown of standard
  space-time structure at small distances, and that this might affect
  the propagation of ordinary particles has led to a program to
  search for violations of Lorentz invariance as a probe of quantum
  gravity.  Initially it was expected that observable macroscopic
  effects caused by microscopic violations of Lorentz invariance would
  necessarily be suppressed by at least one power of the small ratio
  between the Planck length and macroscopic lengths.  Here we discuss
  the implications of the fact that this expectation is in
  contradiction with standard properties of radiative corrections in
  quantum field theories.  In normal field theories, radiative
  corrections in the presence of microscopic Lorentz violation give
  macroscopic Lorentz violation that is suppressed only by the size of
  Standard Model couplings, in clear conflict with observation.  In
  general, this conclusion can only be avoided by extreme fine tuning
  of the parameters of the theory.
\end{abstract}

%=========================================================

\section{Introduction}
\label{sec:introduction}

Although there is enormous uncertainty about the nature of quantum
gravity (QG), one thing is quite certain: The commonly used ideas of
space and time should break down at or before the Planck length is
reached.  For example, elementary scattering processes with a
Planck-sized center-of-mass energy create large enough quantum
fluctuations in the gravitational field that space-time can no longer
be treated as a classical continuum.  It is then natural to question
the exactness of the Lorentz invariance (LI) that is pervasive in all
more macroscopic theories.  Exact LI requires that an object can be
arbitrarily boosted.  Since the corresponding Lorentz contractions
involve arbitrarily small distances, there is an obvious tension with
the expected breakdown of classical space-time at the Planck length.
Indeed, quite general arguments are made that lead to violations of LI
within the two most popular approaches towards QG: string theory
(Ellis {\it et al.}, 2000)
%\cite{string}
and loop quantum gravity (Gambini \& Pullin, 1999; Alfaro {\it et
  al.}, 2000, 2002).
%\cite{ Gambini-Pullin, AUM-T}

This has given added impetus to the established line of research
dedicated to the investigation of ways in which fundamental
symmetries, like LI or CPT, could be broken (Kosteleck\'y {\it et
  al.}, 1989a, 1989b, 1991, 1996).
%cite{kostelecky1},
It was realized that extremely precise tests could be made with a
sensitivity appropriate to certain order of magnitude estimates of
violations of LI (Amelino-Camelia {\it et al.}, 1998).
 %\cite{ACamelia1}.

The sensitivity of the tests arises because there is a universal
maximum speed when LI holds, and even small modifications to the
standard dispersion relation relating energy and 3-momentum give
highly magnified observable effects on the propagation of
ultra-relativistic particles.  One possible modification is
\begin{equation}
 \label{mod.dispersion}
 E^2 = P^2 + m^2 + \frac{ \xi }{ M_{\rm Pl} } E^3.
\end{equation}
Here $E$, and $p$ are a particle's energy and momentum in some
preferred frame, $m$ is its mass, while $\xi$ is a dimensionless
parameter arising from the details of the QG effects on the particular
particle type.  $\xi$ could depend on the particle species and its
polarization.  The dispersion relation can be written in a covariant
fashion:
\begin{equation}
 \label{mod.dispersion.cov}
 P^\mu P_\mu = m^2 + \frac{\xi}{M_{\rm Pl}} (P^\mu W_\mu)^3 ,
\end{equation}
where $P^\mu$ is the particle's four momentum, and $W^\mu$ is the
4-velocity of the preferred frame.  Amelino-Camelia {\it et al.}\ (1998)
%\cite{ACamelia1}
noted that photons ($m=0$) with different energies would then travel
with different velocities.  For a gamma ray burst originating at a
distance $D$ from us, the difference in time of arrival of different
energy components would be $\Delta t =\xi D \Delta E/M_{\rm Pl}$.  If
the parameter $\xi$ were of order $1$ and $D \sim \unit[100]{Mpc}$,
then for $\Delta E \sim \unit[100]{MeV}$, we would have $\Delta t \sim
\unit[10^{-2}]{s}$, making it close to measurable in gamma ray bursts.

A second possible modification is that the parameter normally called
the speed of light, $c$, is different for different kinds of particle.
This is implemented by a non-universal particle-dependent coefficient
of $P^2$ in Eq.\ (\ref{mod.dispersion}).  The differences in the
maximum speeds of propagation also gives sensitive tests: vacuum
Cerenkov radiation etc (Coleman \& Glashow, 1999).

There are in fact two lines of inquiry associated with modified
dispersion relations.  One is the initial approach, where the
equivalence of all reference frames fails, essentially with the
existence of a preferred frame.  A second popular approach preserves
the postulate of the equivalence of all frames, but
tries to find modifications of the standard Lorentz or Poincar\'e
symmetries.  The most popular version, with the name of Doubly Special
Relativity (DSR), replaces the standard Poincar\'e algebra by a
non-linear structure (Amelino-Camelia, 2002; Magueijo \& Smolin, 2002;
Kowalski-Glikman \& Nowak, 2002; Lukierski \& Nowicki, 2003).
 % \cite{DSR}.
Another line of argument examines a deformed algebra formed by
combining the Poincar\'e algebra with coordinate operators one
(Vilela Mendez, 1994; Chryssomalakos \& Okon, 2003, 2004).
% \cite{VMendes}.
Related to these are field theories on non-commutative space-time
 %\cite{NCFT}
(Chaichian {\it et al.}, 2004; Aschieri {\it et al.}, 2005; Douglas \&
Nekrasov, 2001; Szabo, 2003); they give a particular kind of LIV at
short distances that fits into the general field theoretic framework
we will discuss. 

In this article we will concentrate on the first issue, actual
violations of LI.  Regarding DSR and its relatives, we refer the
reader to a contribution in this volume and to critiques by
Sch\"utzhold \& Unruh (2003), by Rembieli\'nski \& Smoli\'nski
(2003), and by Sudarsky (2005).
%\cite{Perspectives, DSRProblems}
A problem that concerns us is that the proposed symmetry algebras all
contain as a subalgebra the standard Poincar\'e algebra, and thus
they contain operators for 4-momentum that obey the standard
properties.  The DSR approach uses a modified 4-momentum that are
non-linear functions of what we regard as the standard momentum
operators. This of course raises the issue of which are the operators directly related to
 observations.  In the discussion section \ref{sec:discussion}, we will
summarize a proposal by Liberati, Sonego and Visser (2004) who propose
that it is the measurement process that picks out the modified
4-momentum operators as the measurable quantities.

We will also touch on an aspect with important connections to the
general field of QG: the problem of a physical regularization and
construction of quantum field theories (QFT).

%=========================================================

\section{Phenomenological models}
\label{sec:models}

Methodical phenomenological explorations can best be quantified
relative to a definite theoretical context.  In our case, of Lorentz
invariance violation (LIV) at accessible energies, the context should
minimally incorporate known microscopic physics, including quantum
mechanics and special relativity (in order to consider small
deviations therefrom).  This leads to the use of a conventional
interacting quantum field theory but with the inclusion of Lorentz
violating terms in the Lagrangian.

One proposal is the Standard Model Extension (SME) of Colladay \&
Kosteleck\'y (1998) and Coleman \& Glashow (1999).  This
incorporates within the Standard Model of particle physics all the
possible renormalizable Lorentz violating terms, while preserving
$\text{SU}(3)\times \text{SU}(2) \times \text{U}(1)$ gauge symmetry and the
standard field 
content.  For example, the terms in the free part of the Lagrangian
density for a free fermion field $\psi$ are:
\begin{align}
{ \cal L}_{\rm free}
 ={}& i\bar\psi ( \gamma_\mu + c_{\mu\nu}\gamma^\nu + d_{\mu\nu}\gamma_5 \gamma^\nu
+e_\mu+i f_\mu \gamma_5 + \tfrac12 g_{\mu \nu \rho}\sigma^{\nu \rho})\partial^\mu \psi
\nonumber\\
& -
\bar\psi(m+a_{\nu}\gamma^\nu + b_{\nu}\gamma_5 \gamma^\nu
+\tfrac12 H_{\nu \rho}\sigma^{\nu \rho})\psi.
\end{align}
Here the quantities $a_\mu$, $b_\mu$, $c_{\mu\nu}$, $d_{\mu\nu}$,
$e_\mu$, $f_\mu$, $g_{\mu \nu \rho}$ and $H_{\mu\nu}$ are numerical
quantities covariantly characterizing LIV, and can be thought of as
arising from the VEV of otherwise dynamical gravitational fields.  The
interacting theory is then obtained in the same way as in the usual,
with $\text{SU}(3)\times \text{SU}(2) \times \text{U}(1)$ gauge fields and a
Higgs field. 
The expected renormalizability was shown by Kosteleck\'y and Mewes
(2001) and Kosteleck\'y {\it et al.}\ (2002).
%\cite{electro}

A second approach, as used by Myers and Pospelov (2003),
% \cite{Myers}
is to take the LIV terms as higher dimension non-renormalizable
operators.  This is a natural proposal if one supposes that LIV is
produced at the Planck scale with power suppressed effects
at low energy; it gives modified dispersion relations at tree
approximation.  For example, there are dimension-5 terms with
$1/M_{\rm Pl}$ suppression in the free part of the Lagrangian, such as
\begin{equation}
\label{eq:dim5.term}
  \frac{1}{M_{\rm Pl}} W^\mu W^\nu W^\rho
  \bar\psi(\xi_f+\xi_{5f}\gamma_5)\gamma_\mu \partial_\nu \partial_\rho
\psi,
\end{equation}
where $W^\mu$ specifies a preferred frame.  Similar terms can be
written for scalar fields and gauge fields.
Dimensionless parameters $\xi$ in these terms specify the degree of
LIV in each sector.

Each of the proposed Lagrangians can be regarded as defining an
effective low-energy theory.  Such a theory systematically provides an
approximation, valid at low energies, to a more exact microscopic
theory.  

In Secs.\ \ref{sec:EffLDTh} and \ref{sec:problems}, we will analyze
the applicability of LIV effective theories.  But first, we will
make some simple model calculations, to illustrate generic features of
the relation between microscopic LIV and low-energy properties of a
QFT.

%=========================================================

\section{Model calculation}
\label{sec:calcs}

The central issue is associated with the UV divergences of
conventional QFT.  Even if the actual divergences are removed because
of the short-distances properties of a true microscopic theory, we
know that QFT gives a good approximation to the true physics up to
energies of at least a few hundred GeV.  So at best the UV divergences
are replaced by large finite values which still leave observable low
energy physics potentially highly sensitive to short-distance
phenomena.

Of course, UV divergences are normally removed by renormalization,
i.e., by adjustment of the parameters of the Lagrangian.  The
observable effects of short-distance physics now appear indirectly,
not only in the values of the renormalized parameters, but also in the
presence in the Lagrangian of all terms necessary for
renormalizability.

The interesting and generic consequences in the presence of Lorentz
violation we now illustrate in a simple Yukawa theory of a scalar field
and a Dirac field.  Before UV regularization the theory is defined by
\begin{align}
\label{eq:L.Yukawa}
{\cal L} ={}& \frac12 (\partial\phi)^2 - \frac{m_0^2}{2} \phi^2
             + \bar\psi (i\gamma^\mu\partial_\mu - M_0) \psi
             + g_0 \phi\bar\psi\psi.
\end{align}
We make the theory finite by introducing a cut-off on spatial momenta
(in a preferred frame defined by a 4-velocity $W^\mu$).  We use a
conventional real-time formalism, so that the cutoff theory is within
the framework of regular quantum theory in 3 space dimensions.  The
cutoff is implemented as a modification of the free propagators:
\begin{align}
\label{propFermions}
    \frac{i}{\gamma^\mu p_\mu -m_0 +i\epsilon}
\to{}&
   \frac{i f(|\ve p|/\Lambda)}
   { \gamma^\mu p_\mu -m_0 +\Delta(|\ve p|/\lambda) +i\epsilon},  
\\
\label{propScalar}
    \frac{i}{p^2 -M_0^2 +i\epsilon}
\to{}&
     \frac{i \tilde f(|\ve p|/\Lambda)}
          {p^2 -M_0^2 +\tilde\Delta(|\ve p|/\lambda) +i\epsilon}.
\end{align}
Here, the functions $f(|\ve p|/\Lambda)$ and $\tilde f(|\ve
p|/\Lambda)$ go to $1$ as $|\ve p|/\Lambda \to 0$, to reproduce normal
low energy behavior, and they go to zero as $|\ve p|/\Lambda\to\infty$,
to provide UV finiteness. The functions $\Delta$ and $\tilde\Delta$
are inspired by concrete proposals for modified dispersion relations,
and they should go to zero when $|\ve p|/\Lambda \to 0$.  But in our
calculations we will set $\Delta$ and $\tilde\Delta$ to exactly zero.
We will assume $\Lambda$ to be of order the Planck scale.

Corrections to the propagation of the scalar field are governed by its
self-energy\footnote{In perturbation theory, the sum over
  one-particle-irreducible two-point graphs. } $\Pi(p)$, which we
evaluate to one-loop order.  We investigate the value when $p^\mu$ and
the physical mass $m$ are much less than the cutoff $\Lambda$.
Without the cutoff, the graph is quadratically divergent, so that
differentiating three times with respect to $p$ gives a convergent
integral (i.e., one for which the limit $\Lambda\to\infty$ exists).  
Therefore we write
\begin{equation}
\label{Pi}
   \Pi(p) =A + p^2 B +
             p^\mu p^\nu W_\mu W_\nu \tilde\xi
   + \Pi^{\rm (LI)}(p^2) + {\cal O}{(p^4/\Lambda^2)} ,  
\end{equation}
in a covariant formalism with
$p^2=p^\mu p^\nu\eta_{\mu\nu}$, where $\eta_{\mu\nu}$ is the
space-time metric.  The would-be divergences at $\Lambda=\infty$ are
contained in the first three terms, quadratic in $p$, so that we can
take the limit $\Lambda\to\infty$ in the fourth term
$\Pi^{\rm(LI)}(p^2)$, which is therefore Lorentz invariant.  The fifth
term is Lorentz violating but power-suppressed.  The coefficients
$A$ and $B$ correspond to the usual Lorentz-invariant mass and wave
function renormalization, and the only unsuppressed Lorentz violation
is in the third term.  Its coefficient $\tilde \xi$ is finite and
independent of $\Lambda$, and explicit calculation (Collins {\it et
al.}, 2004) gives:
\begin{equation}
\label{vani}
 \tilde  \xi = \frac{g^2} {6\pi^2}
       \left[ 1 + 2 \int \limits_0^{\infty}{dx} x f'(x)^2 \right] .
\end{equation}
Although the exact value depends on the details of the function $f$,
it is bounded below by $g^2/6\pi^2$.  Lorentz violation is therefore
of the order of the square of the coupling, rather than
power-suppressed.  The LIV term in (\ref{Pi}) behaves like a
renormalization of the metric tensor and hence of the particle's
limiting velocity.  The renormalization depends on the field and the
size of the coupling, so that we expect different fields in the
Standard Model to have limiting velocities differing by $\sim
10^{-2}$.  The rough expected size depends only on UV power counting
and Standard-Model couplings.

The expected size is in extreme contrast to the measured limits.  To
avoid this, either Lorentz-violation parameters in the microscopic
theory are extremely fine-tuned, or there is a mechanism that
automatically removes low-energy LIV even though it is present
microscopically.  More exact calculations would use renormalization
group methods.  But we know from the running of Standard-Model
couplings, that this can produce changes of one order of magnitude,
not twenty.

We could also perform the same calculation in conventional
renormalization theory.  We would use a Lorentz-invariant UV regulator
followed by renormalization and removal of the regulator.  The results
would be of the same form, except that that coefficients $A$ and $B$
would change in value and $\xi$ would be zero.  If we regard our
theory with the spatial-momentum cutoff as an analog of a true
Lorentz-violating microscopic theory, we deduce that it agrees with
conventional Yukawa theory with suitable values of its parameters
provided only that an explicitly Lorentz violating term proportional
to $(W\cdot\partial\phi)^2$ is added to its Lagrangian.

%=========================================================

\section{Effective long-distance theories}
\label{sec:EffLDTh}

Normally, the details of physical phenomena on very small distance
scales do not directly manifest themselves in physics on much larger
scales.  For example, a meteorologist treats the atmosphere as a
continuous fluid on scales of meters to many kilometers, without
needing to know that the atmosphere is not a continuum but is
made up of molecules.

In a classical field theory or the tree approximation of a QFT, the
transition from a discrete approximation to a continuum is a simple
matter of replacing discrete derivatives by true derivatives, without
change of parameters.  But in QFT, the situation is much less trivial,
and is formalized in the concept of a ``long-distance effective
theory''.  This provides an approximation to a more exact
microscopic theory, and the errors are a power of $l/D$, where $l$ is
the intrinsic distance scale associated with the microscopic theory,
while $D$ is the much larger distance scale of the macroscopic
phenomena under consideration.  

The effective field theory approach has become particularly important
because of the repeated discovery of particles corresponding to fields
with ever higher mass.  To the extent that gravity is ignored so that
we can stay within the framework of QFT, the relation between
effective theories appropriate for different scales has become
extremely well understood (e.g., Rothstein, 2003).  The basic theorems
build from the decoupling theorem of Appelquist and Carazzone (1975).
(See also Weinberg (1996).)

Both the ideas of an effective field theory and the complications when
the microscopic theory is Lorentz violating were illustrated by our
calculation in the previous section.  For phenomena at low energies
relative to some large intrinsic scale $\Lambda$ of a complete theory,
we have agreement, up to power-suppressed corrections, of:
\begin{enumerate}
\item Calculations in the exact microscopic theory.  This theory, as
  concerns quantum gravity, is not yet known.
\item Calculations in a renormalized low-energy continuum field theory
  whose Lagrangian contains only renormalizable terms, i.e., of
  dimension four or less, possibly supplemented by power-suppressed
  higher-dimension non-renormalizable terms.
\end{enumerate}
A basic intuition is obtained by the use of Wilsonian methods, where
the most microscopic degrees of freedom are integrated out.  At the
one-loop level, these give unsuppressed contributions to low-energy
phenomena of a form equivalent to vertices in a renormalizable
Lagrangian, as with the first three terms in Eq.\ (\ref{Pi}).  This
and its generalizations to all orders of QFT show that a renormalized
effective QFT gives the dominant low energy effects of the microscopic
theory.  A renormalizable low-energy effective theory is
self-contained and self-consistent: it contains no direct hints that
it is an approximation to a better theory.  In constructing candidate
approximate theories of physics, we now treat renormalizability not as
an independent postulate but as a theorem.  

In our model calculations, the theory with a cutoff stands in for the
true microscopic theory.  Our calculations and their generalizations
show that the low energy effective theory is an ordinary
renormalizable QFT but with a LIV Lagrangian, just like the Standard
Model Extension.

Higher power corrections, in $p/\Lambda$ can be allowed for by
including higher-dimension non-renormalizable terms in the Lagrangian
of the effective theory, as in Eq.\ (\ref{eq:dim5.term}).  Loop
corrections derived from the non-renormalizable terms involve a series
of counterterm operators in the Lagrangian with ever higher dimension.
But these also correspond to a suppression by more inverse powers of
$\Lambda$, so it is consistent to truncate the series.  The natural
sizes of the coefficients in the Lagrangian are set in the Wilsonian
fashion by integrals in the effective theory with cutoffs of order the
intrinsic scale $\Lambda$ of the full theory.

However, the phenomenological use of non-renormalizable terms does
imply a definite upper limit on the energies where it is appropriate
to use them.  A classic case is the four-fermion form of weak
interactions, where the limit is a few hundred GeV.  The form of the
interaction gave enough hints to enable construction of the full
Standard Model.  The four-fermion interaction (with some additions)
now arises as the low-energy limit of processes with exchange of $W$
and $Z$ bosons.

An issue very important to the treatment of LIV and quantum gravity is
that, normally, the terms in the Lagrangian a low-energy effective
theory must be all those consistent with the unbroken symmetries of
the microscopic theory.  If some of the terms are observed to be
absent, that gives strong implications about the microscopic theory.
A good example is given by QCD.  At short distances, weak interactions
lead to violations of electromagnetic strength of symmetries such as
parity.  But at energies of a few GeV, it is measured that these
symmetries are much more exact; that is why the weak interactions are
called weak.  As Weinberg (1973) showed, a generic unified theory
would not give this weak parity violation.  He then observed that if
the strong-interaction group commutes with the weak-interaction group,
then the unobserved symmetry violation can be removed by a
redefinition of the fields.  This leads essentially uniquely to QCD as
the strong-interaction part of the Standard Model.

In one respect, the situation with gravity is different from the usual
kinds of effective field theory.  Low energy gravitational physics is
described by a non-renormalizable Lagrangian but is not power
suppressed.  The reasons are that the graviton has zero mass and that
macroscopic \emph{classical} gravitational fields occur, with coherent
addition of the sources.  The standard power-law suppression of
gravitation occurs for quantum interactions of small numbers of
elementary particles.    Unsuppressed gravitational phenomena involve
macroscopic classical fields, which need not be treated by quantum
theoretic methods. 

Modulo this qualification, we get the standard result that the total
(leading-power) effect of the microscopic (Plank-scale) physics on
GeV-scale physics is in determining the values of the renormalized
parameters of the theory, and in changing them from the values
obtained from the naive classically motivated considerations.  This
accounts for the folklore that macroscopic manifestations of
Planck-scale physics are to be found only in power-suppressed
phenomena.

However, for our purposes, the folklore is wrong because it ignores
the price of the low-energy effective theory: that its Lagrangian must
contain \emph{all} renormalizable terms consistent with the symmetries
of the \emph{micro}scopic theory.  If Lorentz symmetry is violated by
Planck-scale physics, then we are inexorably led not the
Lorentz-invariant Standard Model, but to its Lorentz-violating
extension.  Observe that because logarithmic divergences are
momentum-independent they are not associated with Lorentz violation.
It is the self-energy (and related graphs) with higher divergences
that are associated with Lorentz violation.  Note that the true
microscopic theory might well be UV finite.  The UV divergences 
concern the ordinary continuum limit for the low-energy effective
theory; their existence is a diagnostic for the presence of
unsuppressed contributions at low energy.

%=========================================================

\section{Difficulties with the phenomenological models}
\label{sec:problems}

The expected sizes of the Lorentz-violating parameters in the models
summarized in Sec.\ \ref{sec:models} raise some serious difficulties,
which we now discuss.  We assume that on appropriate distance scales,
presumably comparable to the Planck length, there is considerable
Lorentz violation.  This is the kind associated with space-time
granularity, and leads in classical theory or tree approximation to
modified dispersion relations like (\ref{mod.dispersion}).

In the case of the SME, which contains only renormalizable terms, the
natural size of the LIV parameters is then that of a one-loop
Standard-Model correction.  Although this appears to have been
recognized by Kosteleck\'y and Potting (1995), the point is quite
obscured in that paper.  The conflict with data means either that
there is also very small Lorentz violation at the Planck scale or that
quantum gravity contains a mechanism for automatically restoring
macroscopic Lorentz invariance.  In either case, it is unjustified to
adhere to the naive expectation that Lorentz violation is expected to
be suppressed by a power of energy divided by $M_{\rm Pl}$, as in
(\ref{mod.dispersion}).

The scheme of Myers and Pospelov (2003) at first appears more natural.
The renormalizable part of their effective low-energy Lagrangian is
the usual Lorentz-invariant one, to which is added a 5-dimensional
operator suppressed by $1/M_{\rm Pl}$ coefficient.  

But as noted by these authors, consistent use of the effective theory
requires that radiative corrections are needed; insertion of a
dimension-5 operator in a self energy generically leads to large
Lorentz-violation from the same power counting as in our model
calculation.  In general it even gives dimension-3 operators enhanced
with a factor of $M_{\rm Pl}$.  They found that they could avoid these
problems by postulating a certain antisymmetry structure for the
tensor coefficient in the dimension-5 operator.

This is still not sufficient.  Consistent use of the theory also
requires iteration of the physical effects that give the dimension-5
operators, and hence, within the effective theory, \emph{multiple}
insertions of these operators.  As shown by Perez and Sudarsky (2003),
this leads back to the LIV dimension-4 operators that one was trying
to avoid.

The overall result is simply a set of particular cases of the general
rule that the terms in the renormalizable part of the Lagrangian are
all those not prohibited by symmetries of the microscopic theory.
Lorentz symmetry is, by the initial hypothesis of all this work, not
among the symmetries.  Starting with Lorentz-violating modifications
of dispersion relations that by themselves are only large at
Planck-scale energies, bringing in virtual loop corrections in QFT
generates integrals over all momenta up to the Planck scale, complete
with the hypothesized Lorentz violation.  This is a direct consequence
of known properties of relativistic QFT, of which the Standard Model
is only one example, and must be obeyed by any theory of quantum
gravity that reproduces known Standard Model physics in  Standard
Model's domain of validity.  Extreme fine tuning of the parameters of
the microscopic theory could be used to evade the conclusion, but this
is generally considered highly inappropriate for a fundamental
microscopic theory of physics.

Thus a very important requirement of a theory of QG is that it should
ensure the absence of the macroscopic manifestation of effects of any
presumed Lorentz-violating microscopic structure of space-time.  This
feature should be robust, without requiring any fine tuning.  Note
that such overriding general considerations have played a critical
role in the discovery of key physical theories in the twentieth
century, from relativity to QCD.  As to experimental data, it can be
seen in retrospect that only a relatively very small set of
experimental data was essential in determining the course of these
developments.

%=========================================================

\section {Direct Searches}

We now give a short account of some of the methods that have yielded
the most important bounds on Lorentz violation.  These experimental
results are important independently of our critiques of their
theoretical motivations.  For a very complete summary of the
situation we refer the reader to the recent review by Mattingly (2005).
%\cite{Mattingly}

In the introduction, we have already mentioned the idea of
Amelino-Camelia {\it et al.}\ (1998) to search for energy-dependent
differences in the times of arrival of gamma rays from gamma bursts.

Another interesting  source of information relies on the expected
parity-violating nature of  some of 
the natural proposals for LIV effects in the propagation of photons
(Gambini \& Pullin, 1999, Myers \& Pospelov, 2003).
%\cite{Gambini-Pullin, Myers}.
This would lead to differences in the propagation
velocity for photons with  different  helicities.
 It was observed that the effects would
 lead to a depolarization of
linearly polarized radiation as it propagates towards the  Earth. Therefore the observation of
linearly polarized radiation from
 distant sources could be used to set important bounds on such effects.
 For instance, Gleiser and Kozameh (2001)
 %\cite{Kozameh-Gleiser}
 found a bound of the order
 $10^{-4}$ for the parameter $\xi$ for the photon.
%using the radiation from  sources at cosmological distances.
 
Another type of bound can be obtained by noting that is quite unlikely
that the Earth would be at rest in the preferred rest frame associated
with the sought-for LIV.  Thus  in an Earth-bound
laboratory Lorentz-violation could appear as violation of the isotropy
of the laws of physics.  Using the prescription for the expected
effects on fermions which arise in the loop quantum gravity scenarios
(Alfaro {\it et al.}, 2000, 2002),
% \cite {AUM-T},
one arrives at an effective SME description.  Measurements rely
on the extreme sensitivity of the Hughes driver type of test of the
isotropy of physics using nuclear magnetic clocks (Chupp {\it et al.},
1989; Bear {\it et al.}, 2000).
 %\cite{ExpSUV}.
The bounds obtained this way are of the order $10^{-5}$ and $10^{-9} $
on parameters that were originally expected to be of order unity.
Then one obtains very stringent bounds on the parameters
characterizing the state of the quantum geometry (Sudarsky {\it et
  al.}, 2002).  Similar constraints can be placed on the effects that
arise in the string theory scenarios (Sudarsky {\it et al.}, 2003).

A further source of severe constraints uses the possibility that
different particle species have different values of their limiting
velocity, as in the SME.  Tests are made by examining the resulting
changes in thresholds and decay properties of common particles.
Coleman and Glashow (1999) obtained a dimensionless bound of
$10^{-23}$ on this kind of Lorentz violation.  Other related arguments
connected to the existence of a bound to the propagation velocity of
particles for modified dispersion relations have been used by
Jacobson {\it et al.}\ (2002, 2003).
% \cite{Jacobson}
These authors noted that the $\unit[100]{MeV}$ synchrotron radiation
from the Crab nebula requires extremely high energy electrons.  They
combined the upper bound on the frequency of synchrotron radiation for
electrons with a given velocity in a given magnetic field with the
fact that there would be an upper bound for any electron's velocity if
$\xi$ for the electron had a particular sign.  In fact the analysis,
carried out within the Myers and Pospelov framework, indicates that at
least for one of the electron's helicities a corresponding $\xi$
parameter, if it had a particular sign, could not have a magnitude
larger than about $10^{-7}$.
 
Finally there is the reported detection of cosmic rays with energies
beyond the GZK cutoff.  We recall that these ultrahigh energy cosmic
rays are thought to be protons whose interaction with the photons of
the cosmic microwave background would prevent them from traveling more
than about $\unit[50]{Mpc}$, while the likely sources are located much
further away. This anomaly is often presented as candidate
observational evidence for LIV (Bird {\it et al.}, 1995; Elbert and
Sommers, 1995; Takeda {\it et al.}, 1998; Abu-Zayyad {\it et al.},
2002; Bergman, 2003; Bahcall \& Waxman, 2003).
%\cite{GZK-LIV}.
Our own feeling is that the list of unexplored alternative
explanations of this anomaly, even if one needs to go beyond
established physics, is much too broad at this time, and thus its
interpretation as a signature of a LIV --- given the difficulties we
discussed here --- is at best premature. Fortunately the
Auger Experiment will become fully operational soon and its results
should help clarify the situation.

%=========================================================

\section{Evading the naturalness argument within QFT}

Several proposals have been made to evade the naturalness problem for
Lorentz violation.

One argument relies essentially on the possibility that a fiducial
symmetry  would protect Lorentz symmetry.  
Jain and Ralston (2005) and
%\cite{Pospelov}
Nibbelink and Pospelov (2005) argue that supersymmetry could be such
symmetry.  At the one-loop level this indeed works: contributions to
self-energy graphs with particles and their superpartners have the
same couplings but opposite signs.  This cancellation is very
reminiscent of the one for the cosmological constant in the same
theories.  However the authors note that, as the Lorentz algebra is a
subalgebra of the supersymmetry algebra, invoking the latter to
protect the former is not entirely consistent.  They then observe that they
would actually need only the translation subalgebra of the
Poincar\'e algebra to be unbroken. However, it is hard to envision a
situation in which a granular space-time would have the full
translation group as a full continuous symmetry. Moreover as it is
well known, even if it is there at some level, supersymmetry must be
broken at low energies.  Then it is difficult to understand how
could it protect the low energy phenomena from the LIV we have been
discussing, while allowing at the same time for violations to be
observable at higher energy scales that are closer to that energy
regime where super-symmetry is presumably unbroken.

Liberati {\it et al.}\ (2005)
%\cite{Liberati-Visser}
treat a condensed matter model of two component Bose-Einstein
condensate as a model system.  LI is associated with monometricity in
the propagation of the two types of quasi-particles. The authors show
that LI can, under certain conditions, be violated at high energies
while being preserved at low energies.  This is achieved by fine
tuning a certain parameter in the model (the interaction with an
external laser source).  The fine tuning is in agreement with our
general results.  

The authors also conclude that their results are a hint that effective
theories in emergent spacetimes could be unreliable beyond the tree
approximation.  Addressed to effective field theories themselves we do
not think that this is correct, since it contradicts the meaning of an
effective field theory.  The Standard Model is an effective field
theory relative to some more complete microscopic theory, and it most
definitely must be used beyond tree approximation with non-trivial UV
renormalization to get its phenomenological successes.  The real
issue, as is apparent from their next sentence, concerns the issue of
the relation between the EFT and the microscopic theory beyond tree
approximation, in the approach of interacting out short distance
degrees of freedom.  However, if the microscopic theory does actually
violate LIV in an essential way at the Planck scale, then a EFT of the
conventional kind derived from Planck-scale consideration will have
LIV operators obeying the usual power counting rules.  The
applicability of EFT is then at all smaller momentum scales, and low
energy phenomena have an expected size of normal one-loop
corrections.  Moreover the microscopic BEC model is a conventional
quantum theory.  

As Liberati, Sonego and Visser (2004) discuss in another paper, which
we will summarize in the discussion section \ref{sec:discussion}, it
is possible that more fundamental issues come into play, perhaps
concerned with measurement in a theory with a dynamical space-time.
These issues would of course make even the principles of the
derivation of an EFT quite different than in normal QFTs.  But they
would also remove the rationale for simple estimates for the sizes of
higher dimension Lorentz-violating operators in an EFT.

Another proposal was made by Alfaro (2005) for a way to generate
naturally small Lorentz violations.  His general idea is to generate
LIV in the integration measure for Feynman graphs.  The proposal
involves two concrete schemes.  One uses a Lorentz violating cutoff
that contains a parameter which when set to zero recovers a Lorentz
invariant situation; the scheme thus has a parametrizably small LIV.
The second scheme involves a Lorentz violating dimensional
regularization scheme, where the standard Minkowski metric
$\eta_{\mu\nu}$ is replaced by
$g_{\mu\nu}=\eta_{\mu\nu}+\alpha\epsilon W_\mu W_\nu$ where $\epsilon
=n-4$ is the small parameter in the dimensional regularization scheme.

In the first scheme the regularization of a one-loop integral is to
modify it by multiplying the integrand by 
\begin{equation}
  R(k) = \frac{-\Lambda^2}{k^2-\Lambda^2+a k_0^2 +i\epsilon},
\end{equation}
where $a=0$ is the Lorentz-invariant case.  This suffers from a
routing dependence and is therefore not well-defined, certainly not as
a complete theory.  Furthermore, in the Lorentz-invariant case $a=0$,
the regulator factor has a pole at $k^2=\Lambda^2$.  This is very
similar to Pauli-Villars regularization, which gives negative metric
states and therefore the regulated theory cannot be considered a
normal quantum theory.  This scheme therefore does not address the
actual situation we are concerned with in quantum gravity.

The second scheme uses dimensional regularization and modifies the
metric in a way that depends on the $\epsilon=0$ pole in the integral
being calculated.  This graph-dependent modification of the metric
does not correspond to any normal definition of a QFT, and no
rationale is given.

%================================================
\section{Cutoffs in QFT and the physical regularization problem}
\label{sec:defining.QFT}

Our results also have important implications for the definition of
QFT.  Given the well-known complications of renormalization, it is
sensible to try defining a QFT as the limit of an ordinary quantum
mechanical theory defined on a lattice of points in real space.  One
could also make time a discrete variable, but this is unnecessary.
Continuum field theory is defined by taking the limit of zero lattice
spacing, with appropriate renormalization of the bare parameters of
the theory.  However, if the cutoff theory is defined on an ordinary
spatial lattice, boost invariance is completely broken by the rest
frame of the lattice.  Therefore all the issues discussed in this
paper apply to the construction of the renormalized continuum limit,
and fine-tuning is needed to get Lorentz invariance.  This is
acceptable for a mathematical definition of a QFT, but not in a theory
that has a claim on being a fundamental theory.

Normal methods of calculation avoid the problem, but in none of them
is the regulated theory a normal quantum mechanical model.  For
instance, the functional integral, as used in lattice gauge theories,
is defined in Euclidean space-time.  The regulated theory on a lattice
is a purely Euclidean construct.  Discrete symmetries under exchange
of coordinate axes are enough to restrict counterterms to those that
give SO(4) invariance in the continuum limit.  Continuum QFT in
Minkowski space is obtained by analytic continuation of the time
variable, and the compact SO(4) symmetry group of the Euclidean
functional integral corresponds to the non-compact Lorentz group in
real space-time.

On the other hand, a Pauli-Villars regulator can preserve LI in the
regulated theory, but only at the expense of negative metric
particles.  That is, the regulated theory is not a normal quantum
mechanical model. 

Finally, dimensional regularization does preserve LI and many other
symmetries.  In this method, space is treated as having a non-integer
dimension.  Technically, space is made infinite dimensional, and this
allows nonstandard definitions to be made of the integrals used in
Feynman graphs so that they behave as if space has an arbitrary
complex dimension (Collins, 1984).
%\cite{JCC.renormalization}.
However, it is not even known how to formulate quantum field
theories non-perturbatively within this framework.

Therefore we pose the problem of whether there exists a physical
regularization of QFT in which LI is preserved naturally.  A physical
regularization means that the regulated theory is a normal quantum
theory whose existence can be taken as assured.

One proposal of this kind was made by Evens {\it et al.}\ (1991),
 %\cite{NonLocalReg}
and it uses a nonlocal regularization.  However Jain and Joglekar
(2004) argue
%\cite{CVNonLocalReg}
 that the scheme
violates  causality and thus is physically unacceptable.

So one is left with a spatial lattice, or some variant, as the only
obvious physical regulator of a QFT.  

The need to treat gravity quantum mechanically provides the known
limits to the physical applicability of the concepts and methods of
QFT.  Therefore the observed Lorentz invariance of real phenomena
indicates that a proper theory of quantum gravity will provide a
naturally Lorentz invariant physical regulator of QFT.  So perhaps a
discovery of a better method of defining a QFT in Minkowski space
might lead to important clues for a theory of QG.

%========================+=================================
\section{Discussion}
\label{sec:discussion}

It is well-known that nontrivial space-time structure is expected at
the Planck scale, and this could easily lead to Lorentz-violating
phenomena.  The simplest considerations suggest that the observable
Lorentz violation is suppressed by at least one power of particle
energy divided by the Planck energy; this small expectation has led to
an ingenious set of sensitive measurements, with so far null results.

However, an examination of field theoretic loop corrections shows that
the expectation is incorrect, in general.  Standard theorems in
quantum field theory show that the low-energy effects of Planck-scale
phenomena can be summarized in an effective low-energy QFT whose
Lagrangian contains all renormalizable terms compatible with the
symmetries of the microscopic theory and the appropriate low-energy
field content; this is the Standard Model Extension.  If there is
Lorentz violation in the fundamental theory, then in the effective
theory, then the Lorentz violating parameters are, as we have shown,
of the size of normal one-loop corrections in the Standard Model, in
violent contradiction with data.  Without some special mechanism,
extreme fine tuning is needed.

It is already known (Susskind, 1979; Weinberg, 1989)
% \cite{fine.tuning}
that there are fine-tuning problems with the Standard Model, involving
at least the cosmological constant, mass hierarchies and the Higgs
mass term.  These, of course, suggest to many physicists that the
Standard Model is not the ultimate microscopic theory, but is a
low-energy approximation to some more exact theory where fine-tuning
is not needed.  Our results show that Lorentz invariance should be
added to the list of fine tuning problems that should be solved by a
good theory that includes quantum gravity, or alternatively by a new
theory that supersedes currently known ideas.  We thus suggest that a
search for a physically meaningful, Minkowskian space-time, Poincar\'e
and gauge invariant regulator for the Standard Model could be
intimately connected with the search for a theory of QG and with its
possible phenomenological manifestations.  The lack of a physical
regularization for QFT besides the lattice makes the non-naturalness
of Lorentz invariance a particularly important problem even when
gravity is left out of the discussion.

We conclude by mentioning some intriguing ideas.

Some ideas regarding how a discrete nature of space-time can be made
consistent with Lorentz invariance are explored by Rovelli \& Speziale
(2003) and by Dowker, Henson, and Sorkin (2004).  In particular Dowker
et al.\ show that by using a random lattice or causal set methods one
can evade the problem that regular spatial lattices prevent a physical
realization of Lorentz contraction.

There are  also considerations  of other possible types of manifestations of QG.  For instance
there are proposals regarding 
nonstandard couplings to the Weyl tensor (Corichi \& Sudarsky, 2005), 
fundamental quantum decoherence (Gambini {\it et al.}, 2004), 
and QG induced collapse of the wave function
(Penrose, 1989; Perez {\it et al.}, 2005).

Finally, there are proposals invoking fundamental modifications of the
Lorentz or Poincar\'e structures.  This is the subject of doubly
special relativity (DSR) which we discussed briefly in our
introduction, Sec.\ \ref{sec:introduction}, together with some
critiques of the physical significance of DSR.  

An interesting idea, with more general applicability, is the proposal by
Liberati, Sonego and Visser (2004) for resolving the problem in DSR
that the measurable momentum operators differ from the operators, also
present in DSR, that obey the standard commutation relations with the
Lorentz generators.  They suggest that the modifications of the  momentum operators
are a non-trivial effect of quantum mechanical measurement when
quantum gravity effects are important.  To our mind, this impinges on
an important foundational problem in QFT and QG as compared with
elementary quantum mechanics, including the issue of the relation
between an effective field theory and an underlying theory in which
space-time is genuinely dynamical.

In simple quantum mechanical theories of systems like the
Schr\"odinger equation for a single atom, measurement involves an
external apparatus.  But with an interacting QFT, the theory is
sufficiently broad in scope that it describes both the system being
measured and the experimental apparatus measuring it.  If the Standard
Model is valid, it accurately governs all strong, electromagnetic and
weak interactions, and therefore it includes particle detectors as
well as particle collisions.  An interacting QFT has a claim on being
a theory of everything (in a certain universe-wide domain) in a way
that a few-body Schr\"odinger equation does not.  Measurement theory
surely has a different status in QFT. This point is exemplified by the
analysis by Sorkin (1993).  This should apply even more so
when quantum gravity is included.  A localized measurement of a
sufficiently elementary particle of sufficiently super-Planck energy
could have a substantial effect on the local space-time metric and
thus on the meaning of the energy being measured.

The emergence of the field known as QG phenomenology is certainly a
welcome development for a discipline long considered as essentially
removed from the empirical realm. However one should avail oneself of
all the other established knowledge in physics, in particular, the
extensive development both at the theoretical and experimental level
of QFT. Ignoring the lessons it provides, and the range of its
successful phenomenology is not a legitimate option, unless one has a
good substitute for it.  The unity of physics demands that we work to
advance in our knowledge by seeking to {\it expand} the range covered
by our theories, therefore we should view with strong skepticism, and
even with alarm any attempt to extrapolate in one direction --- based
essentially on speculation --- at the price of having to cede
established ground in any other.

%=========================================================
\section*{Acknowledgments}

This work was supported in part by the U.S.\ Department of Energy
under grant number DE-FG02-90ER-40577 and by DGAPA-UNAM
IN108103 and CONACyT 43914-F grants (M\'exico).  We would like to thank J.
Banavar, Y. Chen, C. Chryssomalakos,  L. Frankfurt, J. Jain, and  M. Strikman for useful
discussions.

%=========================================================

\end{document}